\documentclass{sig-alternate}
\pdfoutput=1
\DeclareFixedFont{\auacc}{OT1}{phv}{m}{n}{12}   
\DeclareFixedFont{\afacc}{OT1}{phv}{m}{n}{10}   
  \usepackage{algorithm}
  \usepackage[noend]{algorithmic} 
  \usepackage{amsmath,amssymb}\usepackage{pslatex}\usepackage{amstext}

\usepackage{balance}
\usepackage{amsmath,amssymb}
\usepackage{theorem}
\usepackage{multirow}
\usepackage{algorithmic} 
\usepackage[tight]{subfigure}
\usepackage{calc}
\usepackage{amssymb}
\usepackage{amstext}
\usepackage{amsmath}
\usepackage{multicol}
\usepackage{pslatex}

\newcommand{\danielcut}[1]{}%
\newcommand{\kamelcut}[1]{}%

\newcommand{\lxor}{\oplus}
\newcommand{\gt}{GT}
\newcommand{\gc}{GC}
\newtheorem{proposition}{Proposition}  
\def\qed{\relax\ifmmode\hskip2em \Box\else\unskip\nobreak\hskip1em $\Box$\fi}
\newtheorem{corollary}{Corollary}  

\usepackage{ifpdf}
\usepackage{color}
\ifpdf
    \providecommand{\myfig}[1]{#1.pdf}
\else
   
   \providecommand{\myfig}[1]{#1.eps}
\fi
\usepackage{url}

\usepackage{paralist}
\begin{document}

%
\conferenceinfo{DOLAP'07,} {November 9, 2007, Lisboa, Portugal.}
\CopyrightYear{2007}
 \crdata{978-1-59593-827-5/07/0011}


\title{A Comparison of Five Probabilistic View-Size Estimation Techniques in OLAP}
%
%
%
%
%

\numberofauthors{2} 
%
\author{
%
%
\alignauthor
Kamel Aouiche and Daniel Lemire\\
       \affaddr{LICEF, Universit{\afacc\'e} du  Qu{\afacc\'e}bec {\afacc\`a} Montr{\afacc\'e}al}\\
       \affaddr{100 Sherbrooke West}\\
       \affaddr{Montreal, Canada }\\
       \email{kamel.aouiche@gmail.com, lemire@acm.org}
}

\maketitle
\begin{abstract}
A data warehouse cannot materialize all possible views, hence we must estimate
quickly, accurately, and reliably the size of views to determine
the best candidates for materialization.
 Many available techniques for view-size estimation make particular
statistical assumptions and their error can be large. Comparatively, unassuming
probabilistic techniques are slower, but they estimate accurately and
reliability very large view sizes using little memory. We compare five
unassuming hashing-based view-size estimation techniques including Stochastic
Probabilistic Counting and \textsc{LogLog} Probabilistic Counting. Our
experiments  show that only Generalized Counting, {Gibbons-Tirthapura}, and
Adaptive Counting provide universally tight estimates irrespective of the size
of the view; of those, only Adaptive Counting remains constantly fast as we
increase
the memory budget. 
\end{abstract}




\section{Introduction}

View materialization is  one of the most effective technique to improve query
performance of data warehouses. Materialized views are physical structures
which improve data access time by precomputing intermediary results. Typical
OLAP queries consist in selecting and aggregating data with  grouping sets
(\textsc{group by} clauses)~\cite{graycube}.  By precomputing many plausible
groupings, we can avoid slow responses due to aggregates over large tables.
Many queries, such as those
 containing conditions (\textsc{having} clauses) can also be computed faster using
these preaggregates. However, materializing views requires additional storage
space and induces maintenance overhead when refreshing the data warehouse.
Moreover, the number of views is  large: there are $2^d$~views in a
$d$-dimensional data cube lattice~\cite{graycube}. Hence, one of the most
important issues in data warehouse physical design is the selection of the
views to materialize, an NP-hard problem~\cite{gup97sel}. Most heuristics for
this problem depend on view-size estimation.

Some view-size estimation techniques make assumptions about the data
distribution and others are ``unassuming''. A common statistical assumption is
uniformity~\cite{gol98met}, but any skew in the data leads to an overestimate.

Generally, while statistically assuming estimators are computed quickly, the
most expensive step being the random sampling, their error can be large and it
cannot be bounded a priori.  We consider several state-of-the-art statistically
unassuming estimation techniques: Probabilistic
Counting~\cite{flajolet1985pca}, \textsc{LogLog} Probabilistic
Counting~\cite{durand2003lcl}, Adaptive Counting~\cite{cai2005fat}, Generalized
Counting~\cite{BarYossef2002}, and {Gibbons-Tirthapura}~\cite{Gibbons2001}.
While relatively expensive, unassuming estimators tend to provide  good
accuracy and reliability~\cite{aouicheiceis2007}.

To use these techniques, we need to hash rows quickly and our theoretical
bounds require at least pairwise independent hash values. Fortunately, while
there can be several dimensions ($d>10$) in a data cube, the number of
attribute values in each dimension is often small compared to  the available
memory. Hence,  we can hash dimensions separately, store the result in main
memory, and combine these fully independent unidimensional hash values into
3-wise independent multidimensional hash values.

Typically, as we allocate more memory,
our algorithms become more accurate, but also slower.
We are concerned with two different usage scenario. Firstly,
we want rough estimates, with errors
as large as 10\%, as quickly as possible. In such cases, we can use tiny memory budgets (less than 1\,MiB). Secondly, we want highly accurate estimates with errors
less than 1\% or 0.1\%. In these instances, we  use several megabytes of memory.

The main result of this paper is
an exhaustive theoretical and experimental comparisons of a wide
range of unassuming view-size estimation techniques. We also
present practical theoretical results on Generalized Counting, a novel algorithm. Finally, we make some recommendations.

\section{Related Work}\label{sec:RealtedWork}
Sample-based, statistically assuming
estimations are typically fast, but can be inaccurate and can still use a lot
of memory. Indeed,  in the worst-case scenario, the histogram of the sample
might be as large
as the view size we are trying to estimate. 
Moreover, it is difficult to derive
unassuming accuracy bounds since the sample might not be representative and the
model might not be a good fit. However, a sample-based algorithm is expected to
be an order of magnitude faster than an algorithm which processes the entire
data set.
Haas et al.~\cite{haas1995sbe}  estimate the view size from the histogram of a
sample: adaptively, they choose a different estimator based on the skew of the
distribution. Faloutsos et al.~\cite{faloutsos1996msd} obtain results nearly as accurate as Haas
et al., that is, an error of approximately 40\%, but with a simpler
algorithm. 

Stochastic Probabilistic Counting~\cite{flajolet1985pca}, \textsc{LogLog}
Probabilistic Counting (henceforth \textsc{LogLog})~\cite{durand2003lcl} and
Adaptive Counting~\cite{cai2005fat} have been shown to provide very accurate
view-size estimations quickly for very large views, but their estimates assume
we have independent hashing. Because of this assumption, their theoretical
bound may
not hold in practice. 

 Gibbons and Tirthapura~\cite{Gibbons2001} derived
an unassuming bound, for an algorithm we will refer to as
{Gibbons-Tirthapura} or
\gt{}, that only requires pairwise independent hashing.
It has been shown recently that if you have $k$-wise independent hashing for
$k>2$ the theoretically bound can be improved
substantially~\cite{viewsizetechreport}. Bar-Yossef et al.~\cite[Section~2]{BarYossef2002}
presented a new scheme which they described as a generalization of  Probabilistic Counting, assuming only pairwise independent hashing. 
The benefit of
these new schemes is that as  long as the random number generator is
truly  random and the hashed values use enough bits, the theoretical bounds
 have to hold irrespective of the size of
the view or of other factors.  We can be certain
to have high accuracy and reliability, but what about speed?


\section{Estimation by Multifractals}\label{sec:multi}

We implemented the statistically assuming algorithm by
Faloutsos~et~al. based on a multifractal model~\cite{faloutsos1996msd}.
Given a sample, all that is required to learn the multifractal
model is the number of distinct elements in the sample $F_0$, the number of
elements in the sample $N'$, the total number of elements $N$, and the number
of occurrences of the most frequent item in the sample $m_\textrm{max}$. Hence,
a very simple implementation is possible (see
Algorithm~\ref{algo:multifractal}). 
The memory usage of this algorithm is determined by the GROUP BY query on the
sample (line~6): typically, a larger sample will lead to a more important memory
usage.
\begin{algorithm}
 \begin{small}\begin{algorithmic}[1]
\STATE \textbf{INPUT:} Fact table $t$ containing $N$ facts \STATE
\textbf{INPUT:} \textsc{group by} query on dimensions $D_1, D_2, \ldots, D_d$
\STATE \textbf{INPUT:} Sampling ratio $0<p<1$ \STATE \textbf{OUTPUT:} Estimated
size of \textsc{group by} query \STATE Choose a sample in $t'$ of size
$N'=\lfloor pN \rfloor$ \STATE Compute $g$=\textsc{group by}($t'$) \STATE let
$m_{\textrm{max}}$ be the number of occurrences of the most frequent tuple
$x_1,\ldots, x_d$ in $g$ \STATE let $F_0$ be the number of tuples in $g$ \STATE
$k \leftarrow \lceil\log  F_0 \rceil$ \WHILE{$F<F_0$} \STATE $p\leftarrow
(m_\textrm{max}/{N'})^{1/k}$ \STATE $F\leftarrow \sum_{a=0}^k {k\choose a}
(1-(p^{k-a}(1-p)^a)^{N'})$ \STATE $k \leftarrow k+1$ \ENDWHILE \STATE
$p\leftarrow (m_\textrm{max}/N)^{1/k}$ \STATE \textbf{RETURN: $\sum_{a=0}^k
{k\choose a} (1-(p^{k-a}(1-p)^a)^N)$}
 \end{algorithmic}
\end{small}
\caption{\label{algo:multifractal}View-size estimation using a multifractal
distribution model.}
\end{algorithm}

\section{Unassuming  Estimation}

\begin{figure}
\centering\includegraphics[width=0.7\columnwidth]{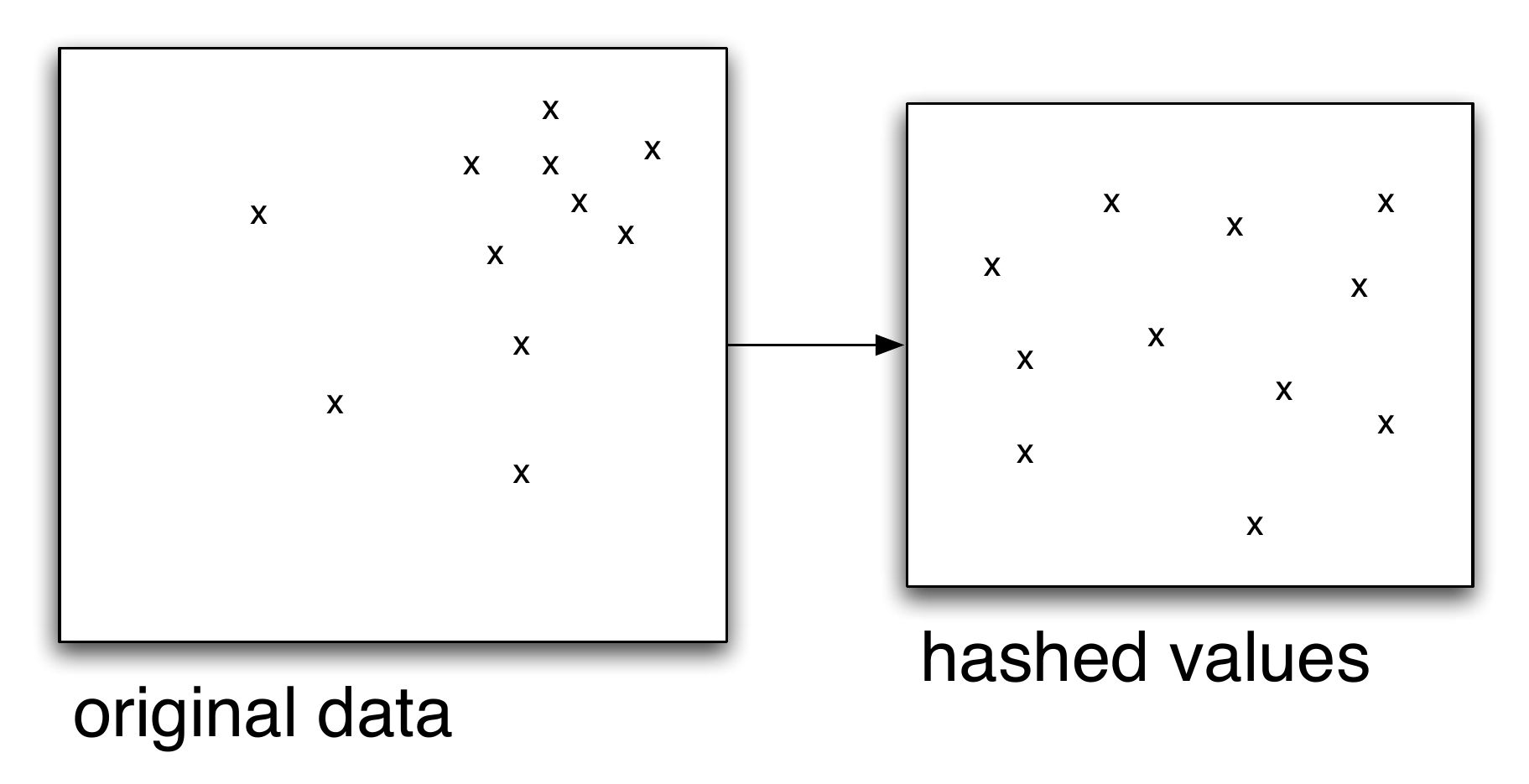}
\caption{\label{fig:hashingtrick}Irrespective of the original data, the hashed values
can be uniformly distributed.}
\end{figure}

All unassuming methods presented in this paper use the same
probabilistic idea. Whereas the initial data has unknown distribution,
if we use an appropriate random hashing method, the hashed values are uniformly distributed (see Fig.~\ref{fig:hashingtrick}).

\subsection{Independent Hashing}\label{sec:hashing}

Hashing maps objects to values in a nearly random way. We are interested in
hashing functions from tuples to $[0,2^L)$ where $L$ is fixed ($L=32$ or $L=64$
in this paper).  Hashing is uniform if $P(h(x)=y)=1/2^L$ for all $x,y$, that
is, if all hashed values are equally likely. Hashing is \textit{pairwise
independent} if $P(h(x_1)=y_1 \land h(x_2)=y_2)= P(h(x_1)=y_1)
P(h(x_2)=y_2)=1/4^L$ for all $x_i,y_i$. Pairwise independence implies
uniformity. Hashing is $k$-wise independent if $P(h(x_1)=y_1 \land \cdots \land
h(x_k)=y_k)=1/2^{kL}$ for all $x_i,y_i$. Finally, hashing is (fully)
independent if it is $k$-wise independent for all $k$.
Fully independent hashing
of $F_0$ distinct values requires $\Omega(F_0)$ units of memory~\cite{alon1986} and is thus impractical if $F_0$ is large.

We can compute 3-wise independent hash values efficiently
 in a multidimensional data warehouse setting. For each
dimension $D_i$, we build a look-up table $T_{i}$, using the attribute values of
$D_i$ as keys.
 Each time we meet a new key,
we generate a random number in $[0, 2^L)$ and store it in the look-up table
$T_i$. This random number is the hashed value of this key.
 This table generates (fully)
independent hash values in amortized constant time. 
In a data warehousing context, whereas dimensions are numerous, each dimension
will typically have few distinct values: for example, there are only 8,760
hours in a year. Therefore, the look-up table will often use a few Mib or less.
When hashing a tuple $x_1,x_2,\ldots,x_k$ in $D_1\times D_2\times \ldots D_k$,
we use the value $T_1(x_1) \lxor T_2(x_2) \lxor \cdots \lxor T_k(x_k)$ where $
\lxor$ is the \textsc{exclusive or} operator. This hashing
 is 3-wise independent and requires amortized constant time.
 Tables $T_i$ can be reused for several estimations: we can simultaneously estimate
the size of a \textsc{group by} on $D_1$ and $D_2$, and the size of a  \textsc{group by} on $D_2$
and $D_3$ while using a single table $T_2$.

\subsection{Probabilistic Counting}\label{sec:probacounting}

\begin{figure}
\centering\includegraphics[width=0.7\columnwidth]{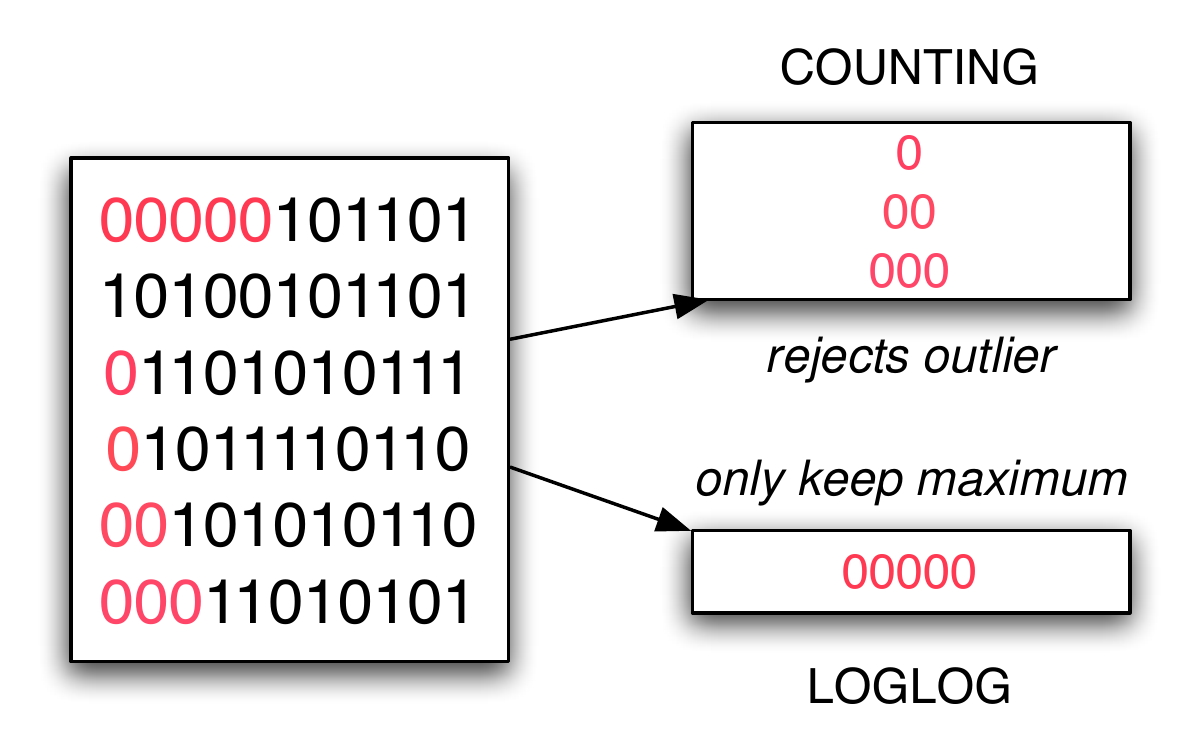}
\caption{\label{fig:countingmethods}Probabilistic counting methods.}
\end{figure}

\begin{figure}[htb]
\begin{center}
\includegraphics[width=0.8\columnwidth]{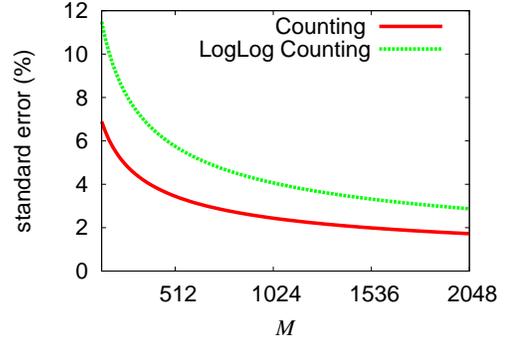}
\end{center}
\caption{ \label{theorycounting}Standard error for Probabilistic Counting and
\textsc{LogLog}  as a function of the memory parameter $M$.
}
\end{figure}

Our version of (Stochastic) Probabilistic
Counting~\cite{flajolet1985pca} (or just Counting for short) is given in  Algorithm~\ref{algo:stoch}.
\textsc{LogLog} (see
Algorithm~\ref{algo:loglog}) is a faster variant~\cite{durand2003lcl}.
 The
main difference between the two algorithms is that \textsc{LogLog} only keeps
track of the maximum number of leading zeroes, whereas Probabilistic Counting
keeps track of all observed numbers of leading zeroes and is thus more
resilient to outliers in the hashing values (see
Fig.~\ref{fig:countingmethods}). For the same parameter $M$, the memory usage
of the two algorithms is comparable in practice: Probabilistic Counting uses a
$M\times L$ binary matrix and \textsc{LogLog} uses $M$ counters to store
integer values ranging from $1$ to $L-\log M$. Assuming independent hashing,
these algorithms have (relative) standard error (or the relative standard
deviation of the error) of $0.78/\sqrt{M}$ and $1.3/\sqrt{M}$ respectively (see
Fig.~\ref{theorycounting}). These theoretical results assume independent
hashing which we cannot realistically provide. They also require the view size
to be very large. 
Fortunately, we can detect the small views.
A small view
 compared to the available memory ($M$),
will leave several of the $M$ counters
 unused (array $\mathcal{M}$ in Algorithm~\ref{algo:stoch}).
Thus, following Cai et al.~\cite{cai2005fat}, when more
than 5\% of the counters are unused
we return a linear counting estimate~\cite{whang1990lin} instead
of the  \textsc{LogLog} estimate: see last line of Algorithm~\ref{algo:stoch} (henceforth Adaptive Counting).
Finally, Alon et al.~\cite{237823} 
presented a probabilistic counting variant using only pairwise independent
hashing, but the error bounds are large: for any $c>2$, the relative error is
bounded by $c-1$ with reliability $1-2/c$ (an error bound of 3900\% 19 times
out of 20). We do not expect these algorithms to be very sensitive to the size of
the memory $M$.

\begin{algorithm}[htb]
\begin{small} \begin{algorithmic}[1]
\STATE \textbf{INPUT:} Fact table $t$ containing $N$ facts
\STATE \textbf{INPUT:} \textsc{group by} query on dimensions $D_1, D_2, \ldots, D_d$
\STATE \textbf{INPUT:} Memory budget parameter $M=2^k$
\STATE \textbf{INPUT:} Independent hash function $h$ from $d$ tuples to $[0,2^L)$.
\STATE \textbf{OUTPUT:} Estimated size of \textsc{group by} query
\STATE $b\leftarrow$ $M\times L$ matrix (initialized at zero)
\FOR{tuple $x\in t$}
  \STATE $x'\leftarrow \pi_{D_1, D_2, \ldots, D_d} (x)$ \COMMENT{projection of the
tuple}
  \STATE $y\leftarrow h(x')$ \COMMENT{hash $x'$ to $[0,2^L)$}
  \STATE $\alpha = y \bmod{M}$ \STATE $i\leftarrow$ position of the first 1-bit in
$\lfloor y/M\rfloor$
  \STATE $b_{\alpha,i}\leftarrow 1$
\ENDFOR
\STATE $A \leftarrow 0$
\FOR{$\alpha \in \{0,1,\ldots, M-1\}$}
  \STATE increment $A$ by the position of the first zero-bit in $b_{\alpha,0},b_{\alpha,1},\ldots$
\ENDFOR
\STATE \textbf{RETURN:} $M/\phi 2^{A/M}$ where $\phi\approx  0.77351$
 \end{algorithmic}\end{small}

\caption{\label{algo:stoch}View-size estimation using
Probabilistic Counting.}
\end{algorithm}

\begin{algorithm}[htb]
\begin{small} \begin{algorithmic}[1]
\STATE \textbf{INPUT:} fact table $t$ containing $N$ facts
\STATE \textbf{INPUT:} \textsc{group by} query on dimensions $D_1, D_2, \ldots, D_d$
\STATE \textbf{INPUT:} Memory budget parameter $M=2^k$
\STATE \textbf{INPUT:}
Independent hash function $h$ from $d$ tuples to $[0,2^L)$.
\STATE \textbf{OUTPUT:} Estimated size of \textsc{group by} query
\STATE $\mathcal{M}\leftarrow \underbrace{0,0,\ldots,0}_M$
\FOR{tuple $x\in t$}
 \STATE
$x'\leftarrow \pi_{D_1, D_2, \ldots, D_d} (x)$ \COMMENT{projection of the
tuple}
 \STATE $y\leftarrow h(x')$ \COMMENT{hash $x'$ to $[0,2^L)$}
 \STATE $j\leftarrow$ value of the first $k$ bits of $y$ in base 2
 \STATE $z\leftarrow$ position of the first 1-bit in the remaining $L-k$ bits of $y$ (count starts at
1)
 \STATE $\mathcal{M}_j \leftarrow \max (\mathcal{M}_j,z)$
\ENDFOR
\STATE (original \textsc{LogLog}) \textbf{RETURN:} $\alpha_M M 2^{\frac{1}{M}\sum_j \mathcal{M}_j}$ \\ where
$\alpha_M  \approx 0.39701-(2\pi^2+\ln^2 2)/(48M)$.
\STATE (Adaptive Counting) \textbf{RETURN:}
$\begin{cases}\alpha_M M 2^{\frac{1}{M}\sum_j \mathcal{M}_j}  & \text{if $ \beta/M \geq 0.051$}\\
-M \log \beta/M & \text{otherwise}
\end{cases}$\\
where $\beta$ is the number of  $\mathcal{M}_j$ for $j=1,\ldots,M$ with value zero
 \end{algorithmic}\end{small}

\caption{\label{algo:loglog}View-size estimation using \textsc{LogLog} and Adaptive Counting.}
\end{algorithm}

\subsection{Generalized Counting}

We modified a generalization to Probabilistic
Counting~\cite[Section~2]{BarYossef2002}  (henceforth \gc{}),
 see Algorithm~\ref{algo:gencounting}.
 The tuples and hash values are stored in an ordered set, and since
each tuple is inserted (line~14), the complexity of processing each tuple with respect
to $M$ is in $O(\log M)$. However, for small $M$ with respect to the view
size, most
tuples are never inserted since their hash value is larger than the smallest $M$~hash values (line~13). 

\begin{figure}
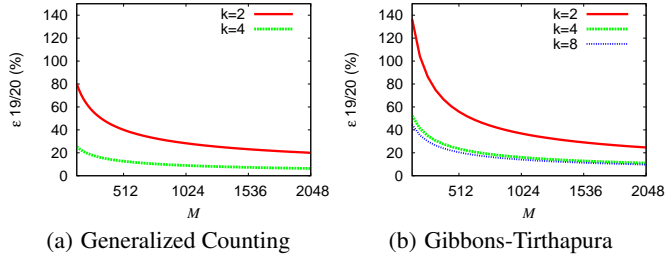

%
\subfigure[\label{gencountingtheory}Generalized Counting]{\includegraphics[height=0.37\columnwidth]{\myfig{gencounting-epsilon-vs-M}}}%
\subfigure[\label{theory}{Gibbons-Tirthapura}
]{\includegraphics[height=0.37\columnwidth]{\myfig{epsilon-vs-M}}}
\caption{Bound on the estimation error (19 times out of 20) as a function of
the number of tuples kept in memory ($M$)  with $k$-wise independent hashing. }
\end{figure}

The original algorithm~\cite{BarYossef2002}
used  many hashing bits: $L \geq 3 \sum_i \log  \vert D_i \vert$ where $\vert D_i \vert$ is the number of attribute values in dimension $D_i$. The main problem is that the number of required bits depends
on the volume of the cuboid, which is typically far larger than
the view-size.
 As the next result shows, with our modified algorithm, few bits
are necessary. For example,
 when hashing with $L=64$~bits, using a memory budget of $M=10000$,
 and with relative accuracy of $\epsilon=0.1$, we can estimate view sizes
 far exceeding anything seen in practice ($2\times 10^{21}$~facts).
 Moreover, we show that the accuracy bounds improve substantially  if the hashed values are more than pairwise independent (see Fig.~\ref{gencountingtheory}).

 \begin{proposition}\label{prop:gencounting}
 For $L \geq 1+ \log F_0/(\epsilon M)$
 and $M \geq 2k \geq 4$,
 Algorithm~\ref{algo:gencounting} estimates a view size $F_0$
within relative precision $\epsilon < 1/2$ with reliability
$1-\delta$ where $\delta$ is given by $ ( 4 k/(e^{2/3} \epsilon^2 M)  )^{k/2}$.
 \end{proposition}
 \begin{proof} Suppose we have $F_0$ distinct tuples in the \textsc{group by}
 and assume that $F_0>M$. If $M\leq F_0$, we can modify
 the algorithm so that an exact count is returned.

 First, consider the case where we overestimate the true count by $\epsilon$, that is
 $2^L M/\textrm{max}(\mathcal{M}) \geq (1+\epsilon) F_0$, hence we have at least
 $M$ hashed values smaller than $2^L M / ((1+\epsilon) F_0 ) $. Hashed values
 take integer values in $[0,2^L)$. Assuming $L\geq 1+ \log F_0/(\epsilon M)$, the probability
 that a hashed value is smaller than $2^L M / ((1+\epsilon) F_0 ) $ is less than $M / ((1+\epsilon) F_0 )+2^{-L}\leq
 M / ((1+\epsilon) F_0 )+\epsilon M/(2 F_0)
 \leq M (2+\epsilon+\epsilon^2)/ (2(1+\epsilon) F_0 ) = M p/F_0$ where $p=  (2+\epsilon+\epsilon^2)/ (2(1+\epsilon)  )$. Let $X_i$ for $i=1,\ldots,F_0$ be 1 with probability
 $p/F_0$ and zero otherwise. Write $X=\sum_{i=1,\ldots,F_0} X_i$, we have that
 $\bar X = \sum_{i=1,\ldots,F_0} E(X_i)= M p$ whereas, by pairwise
 independence,
 $\sigma^2 = var(X) =\sum_{i=1,\ldots,F_0} var(X_i) =
 F_0 (Mp/F_0-M^2 p^2/F_0^2) = Mp (1- Mp/F_0) \leq  Mp.$
 By a Chernoff-Hoeffding bound~\cite[Theorem 2.4]{schmidt1993chb} and the $k$-wise independence of the $X_i$'s,
 $P(X\geq M )\leq
 P(|X-Mp|>M-Mp)
 \leq  \left ( \frac{k M p}{e^{2/3} (1-p)^2 M^2} \right )^{k/2}
 = \left ( \frac{k p}{e^{2/3} (1-p)^2 M} \right )^{k/2}$.
 We have that $p\leq 1$ and $1-p\geq \epsilon /2 $ for $\epsilon < 1/2$,
 hence $P(X\geq M )\leq \left ( \frac{k 4}{e^{2/3} \epsilon^2 M} \right )^{k/2}$. Finally, observe that $P(2^L M/\textrm{max}(\mathcal{M}) \geq (1+\epsilon) F_0) \leq P(X\geq M)$.

Similarly,
suppose that we underestimate the true count by $\epsilon$,
 $2^L M/\textrm{max}(\mathcal{M}) \leq (1-\epsilon) F_0$, hence we have less than
 $M$ hashed values smaller than $2^L M / ((1-\epsilon) F_0 ) $.
 The probability
 that a hashed value is smaller than $2^L M / ((1-\epsilon) F_0 ) $ is less than $M / ((1-\epsilon) F_0 )+2^{-L}\leq
 M / ((1-\epsilon) F_0 )+\epsilon M/(2 F_0)
 \leq M (2+\epsilon-\epsilon^2)/ (2(1-\epsilon) F_0 ) = M p/F_0$ where $p=  (2+\epsilon-\epsilon^2)/ (2(1-\epsilon)  )$.
  Let $X_i$ for $i=1,\ldots,F_0$ be 1 with probability
 $p/F_0$ and zero otherwise. Write $X=\sum_{i=1,\ldots,F_0} X_i$, we have that
 $\bar X =  M p$ whereas $\sigma^2 = Mp$. Finally,
 $P(X\leq M )\leq  P(|X-Mp|>M-Mp)\leq \left ( \frac{k p}{e^{2/3} (1-p)^2 M} \right )^{k/2}$. By inspection, we see  that $p/(1-p)^2 \leq 2/\epsilon^2$,
 hence $P(X\leq M )\leq \left ( \frac{k 4}{e^{2/3} \epsilon^2 M} \right )^{k/2}$ which completes the proof.
 \end{proof}

\begin{algorithm}[t]
\begin{small} \begin{algorithmic}[1]
\STATE \textbf{INPUT:} Fact table $t$ containing $N$ facts
\STATE \textbf{INPUT:} \textsc{group by} query on dimensions $D_1, D_2, \ldots, D_d$
\STATE \textbf{INPUT:} Memory budget parameter $M$
\STATE \textbf{INPUT:} $k$-wise  hash function $h$ from $d$ tuples to $[0,2^L)$. \STATE \textbf{OUTPUT:} Estimated size of \textsc{group by} query
\STATE $\mathcal{M}\leftarrow$ empty sorted sequence,  $\textrm{max}(\mathcal{M})$ returns an element with largest hashed value
\STATE $t\leftarrow0$
\FOR{tuple $x\in t$}
\STATE $x'\leftarrow \pi_{D_1, D_2, \ldots, D_d} (x)$ \COMMENT{projection of the tuple}
\STATE $y\leftarrow h(x')$ \COMMENT{hash $x'$ to $[0,2^L)$}
\IF{size($\mathcal{M}$)$< M$}
\STATE insert $x'$ with hashed value $y$ in $\mathcal{M}$
\ELSIF{$y<\textrm{max}(\mathcal{M})$}\label{condline}
\STATE insert $x'$ with hashed value $y$ in $\mathcal{M}$\label{insertline}\COMMENT{$x'$ may already be in $\mathcal{M}$}
\IF{size($\mathcal{M}$)$> M$}
\STATE remove $\textrm{max}(\mathcal{M})$ from $\mathcal{M}$
\ENDIF
\ENDIF
\ENDFOR
\STATE \textbf{RETURN:} $2^L \textrm{size}(\mathcal{M})/\textrm{max}(\mathcal{M})$ 
 \end{algorithmic}\end{small}
\caption{\label{algo:gencounting}Generalized Counting view-size estimation.}
\end{algorithm}

\subsection{Gibbons-Tirthapura}\label{sec:gibbons}

Originally, the \gt{} algorithm was proposed in
the context of data streams and parallel processing~\cite{Gibbons2001} (see
Algorithm~\ref{algo:gibbons}).  If the view
size is smaller than the memory parameter ($M$), the
 estimation is
without error. For this reason, we expect \gt{} to
perform well when estimating
small and moderate view sizes compared to the available memory. 
We can processing most tuples  in (amortized)
constant time with respect to $M$ (line~13) using a hash table,
however the occasional pruning of tuples requires (amortized) linear time
with respect to $M$ (line~16).

\begin{algorithm}[t]
\begin{small} \begin{algorithmic}[1]
\STATE \textbf{INPUT:} Fact table $t$ containing $N$ facts \STATE
\textbf{INPUT:} \textsc{group by} query on dimensions $D_1, D_2, \ldots, D_d$
\STATE \textbf{INPUT:} Memory budget parameter $M$
\STATE \textbf{INPUT:} $k$-wise  hash function $h$ from $d$ tuples to $[0,2^L)$.
\STATE \textbf{OUTPUT:} Estimated size of \textsc{group by} query
\STATE $\mathcal{M}\leftarrow$ empty look-up table
\STATE $t\leftarrow0$ \FOR{tuple $x\in t$}
\STATE $x'\leftarrow \pi_{D_1, D_2, \ldots, D_d} (x)$
\COMMENT{projection of the tuple}
\STATE $y\leftarrow h(x')$ \COMMENT{hash $x'$
to $[0,2^L)$}
\STATE $j\leftarrow$ position of the first 1-bit in $y$ (count
starts at 0)
\IF{$j \leq t$}
\STATE $\mathcal{M}_{x'}=j$\label{linesetting}
\WHILE{$\textrm{size}(\mathcal{M})>M$}
\STATE $t\leftarrow t+1$
\STATE prune
all entries in $\mathcal{M}$ having value less than $t$ \label{linepruning}
\ENDWHILE \ENDIF
\ENDFOR \STATE \textbf{RETURN:} $2^t \textrm{size}(\mathcal{M})$
 \end{algorithmic}\end{small}
\caption{\label{algo:gibbons}{Gibbons-Tirthapura} view-size estimation.}
\end{algorithm}


The original theoretical bounds~\cite{Gibbons2001} assumed pairwise
independence.  However, more independent hashing, as is possible in our context
for views with many dimensions, allow for better theoretical
bounds~\cite{viewsizetechreport} as
illustrated by Fig.~\ref{theory}. Comparing Fig.~\ref{theory}
and~\ref{gencountingtheory}, we may be tempted to conclude
that \gc{} is far superior to \gt{}. We will compare them experimentally.

\begin{proposition}\label{countprop}Algorithm~\ref{algo:gibbons} estimates the number of distinct tuples
within relative precision $\epsilon$, with a  $k$-wise independent hash for
$k\geq 2$ by storing $M$ distinct tuples ($M\geq 8 k$) and  with reliability
$1-\delta$ where $\delta$ is given by
\newcommand{\frack}[2]{{{#1}/{#2}}}  
\begin{eqnarray*}\delta
  &\leq & \frac{k^\frack{k}{2}}{ e^{\frack{k}{3}}  M^{\frack{k}{2}}  } \left ( \frac{\alpha^{\frack{k}{2}}}{(1-\alpha)^k}+
\frac{4^{\frack{k}{2}}}{\alpha^{\frack{k}{2}} \epsilon^k
(2^{\frack{k}{2}}-1)}\right ) .
\end{eqnarray*}
for $4k/M\leq \alpha<1$ and any $k,M>0$.
\end{proposition}

For the case where hashing is 4-wise independent, we derived a more concise bound~\cite{aouicheiceis2007}.
\begin{corollary}\label{corollary1}
 With 4-wise independent hashing, Algorithm~\ref{algo:gibbons} estimates
 the number of distinct tuples
within relative precision $\epsilon\approx 5/\sqrt{M} $, 19 times out of 20 for
$\epsilon$ small.
\end{corollary}

\danielcut{
\begin{proof}\newcommand{\frack}[2]{{{#1}/{#2}}}  
 We start from the 
 inequality of Proposition~\ref{countprop}.
Differentiating
$\frac{\alpha^{\frack{k}{2}}}{(1-\alpha)^k}+\frac{4^{\frack{k}{2}}}{\alpha^{\frack{k}{2}}
\epsilon^k (2^{\frack{k}{2}}-1)}$ with respect to $\alpha$ and setting the
result to zero, we get $3\alpha^4 \epsilon^4+16 \alpha^3-48 \alpha^2-16=0$
(recall that $4k/M\leq \alpha<1$). By multiscale analysis, we seek a solution
of the form $\alpha = 1-a\epsilon^r+o(\epsilon^r)$ and we have that
$\alpha\approx 1- \frack{1}{2}\sqrt[3]{\frack{3}{2}} \epsilon^{4/3}$ for
$\epsilon$ small. Substituting this value of $\alpha$, we have
$\frac{\alpha^{\frack{k}{2}}}{(1-\alpha)^k}+
\frac{4^{\frack{k}{2}}}{\alpha^{\frack{k}{2}} \epsilon^k
(2^{\frack{k}{2}}-1)}\approx \frac{128}{24 \epsilon^4}$. The result follows by
substituting in the second inequality.
\end{proof}
}

\section{Experimental Results}\label{sec:Experiment}

To benchmark the accuracy and speed of our implementation of the view-size
estimation algorithms, we have run tests over the US~Census~1990 data
set~\cite{KDDRepository} as well as on synthetic data produced by
DBGEN~\cite{DBGEN}. The synthetic data was produced by running the DBGEN
application with scale factor parameter equal to 2 except where otherwise
stated. The characteristics of data sets are detailed in
Table~\ref{tab:caractDataSet}. We selected 20 and 8 views respectively from
these data sets: all views in US~Census~1990 have at least 4~dimensions whereas
only 2 views have at least 4~dimensions in the synthetic data set.
Statisticians sometimes define the standard error to be the standard deviation
of the measures, but when the exact value can be known, it is better to use the
deviation from the true value or $\sqrt{E((X-c)^2)}/c$ where $c$ is the value
we try to estimate. The (relative) standard error, defined as the standard
deviation of the error, was computed from 20~estimates using this formula where
$c$, the exact count, was computed once using brute force.

\begin{table}[!h]
    \centering
    \begin{scriptsize}
    \begin{tabular}{ccc} \hline
     & \textbf{US~Census~1990} & \textbf{DBGEN} \\ \hline
    \# of facts& 2458285 & 13977981 \\
    \# of views& 20 & 8 \\
    \# of attributes& 69 & 16 \\ \hline
    Data size& 360\,MiB & 1.5\,GiB\\ \hline
    \end{tabular}
    \end{scriptsize}
    \caption{Characteristic of data sets.}\label{tab:caractDataSet}
\end{table}

We used the GNU C++ compiler version~4.0.2 with the ``-O2'' optimization flag
on an Apple MacPro machine with 2~Dual-Core Intel Xeon processors running at
2.66\,GHz  and 2\,GiB of RAM. No thrashing was observed. To ensure
reproducibility, C++ source code is available freely at
\url{http://code.google.com/p/viewsizeestimation/}.  For the US~Census~1990
data set, the hashing look-up table is a simple array since there are always
fewer than 100~attribute values per dimension. Otherwise, for the synthetic
DBGEN data, we used the GNU/CGI STL extension \texttt{hash\_map} which is to be
integrated in the C++ standard as an \texttt{unordered\_map}: it provides
amortized $O(1)$ inserts and queries. All other look-up tables are implemented
using the STL \texttt{map} template which has the computational complexity
 of a red-black tree. We used comma separated (CSV) (and pipe
separated files for DBGEN) text files and wrote our own C++ parsing code.
%

The test protocol we adopted 
    (see Algorithm~\ref{algo:protocol})
    has been
executed for each unassuming estimation technique
, \textsc{group by} query, random seed and memory size. At each step
corresponding to those parameter values, we compute the estimated \textsc{group
by} view sizes and time required for their computation. Similarly, for the
multifractal estimation technique, we computed the time and estimated size for
each \textsc{group by}, sampling ratio value and random seed.
\begin{algorithm}[t]
\begin{small}\begin{algorithmic}[1]





\FOR{\textsc{group by} query $q\in Q$}
    \FOR{memory budget $m\in M$}

        \FOR{random seed value $r\in R$}

            \STATE Estimate the size of \textsc{group by} $q$ with $m$ memory budget and $r$
            random seed value
            \STATE Save estimation results (time and estimated size) in a log
            file
        \ENDFOR
    \ENDFOR
\ENDFOR
\end{algorithmic}\end{small}
\caption{\label{algo:protocol}Test protocol.}
\end{algorithm}

In Subsection~\ref{sec:smallmemory}, we consider the first use case: the user
is satisfied with a moderate accuracy (such as 10\%). In
Subsection~\ref{sublarge}, we address the case where high accuracy (at least
1\%) is sought, maybe at the expense of memory usage and processing speed.

\subsection{Small memory budgets}\label{sec:smallmemory}

\subsubsection{Accuracy}

\subsubsection*{Test over the US Census 1990 data set} \label{sub:smallcensus}

\begin{figure*}[!t]
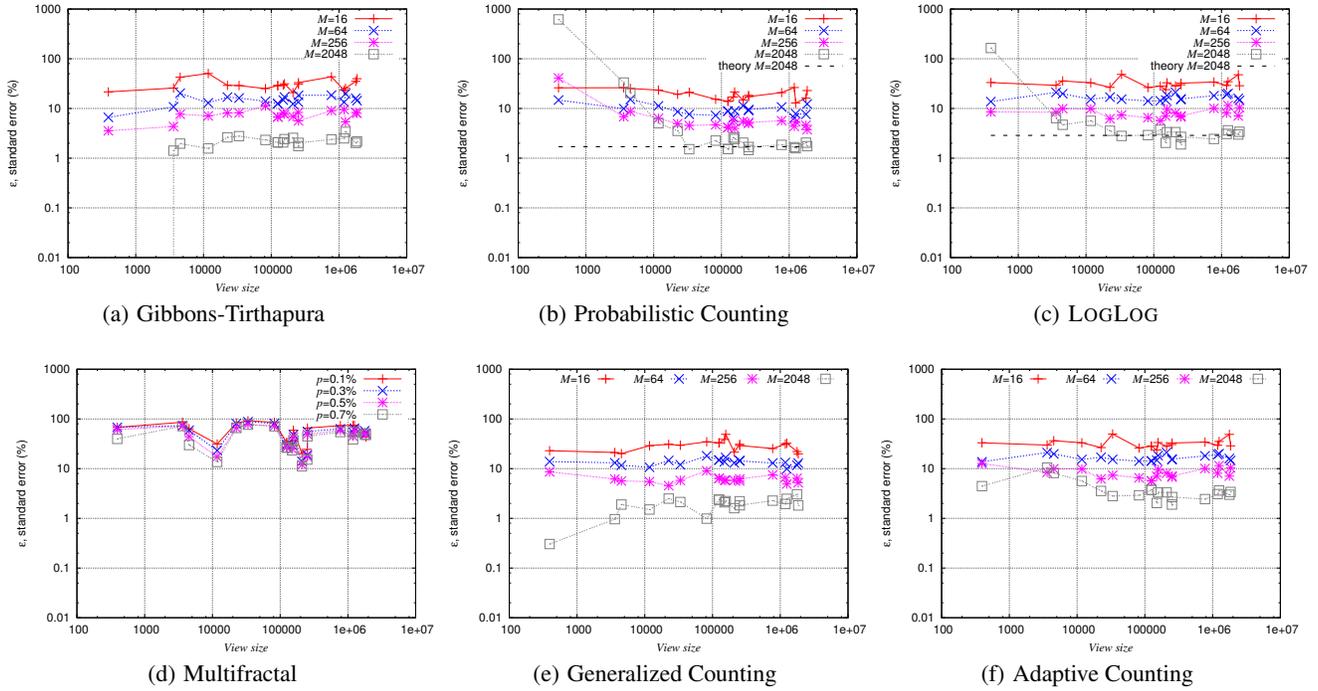

\begin{center}
  \subfigure[{Gibbons-Tirthapura}]{\includegraphics[width=0.32\textwidth]{\myfig{gibbons_error}}\label{fig:expUscensusErrorGibbons}}\quad
  \subfigure[Probabilistic Counting]{\includegraphics[width=0.32\textwidth]{\myfig{counting_error}}\label{fig:expUscensusErrorCounting}}
  \subfigure[\textsc{LogLog}]{\includegraphics[width=0.32\textwidth]{\myfig{loglog_error}}\label{fig:expUscensusErrorLogLog}}\quad
  \subfigure[Multifractal]{\includegraphics[width=0.32\textwidth]{\myfig{multifractal_error}}\label{fig:expUscensusErrorMulti}}
  \subfigure[Generalized Counting]{\includegraphics[width=0.32\textwidth]{\myfig{baryossef_error}}\label{fig:expUscensusErrorBar}}
  \subfigure[Adaptive Counting]{\includegraphics[width=0.32\textwidth]{\myfig{adaptive_error}}\label{fig:expUscensusErrorAdaptive}}
\end{center}
\caption{Standard error of estimation as a function of exact view size for
increasing values of $M$ (US Census 1990).} \label{fig:expUscensusError}
\end{figure*}

Fig.~\ref{fig:expUscensusError} represents the standard error for each
unassuming estimation technique and memory size $M \in \{16, 64, 256, 2048\}$.
For the multifractal estimation technique, we present the standard error for
each sampling ratio $p \in \{0.1\%, 0.3\%, 0.5\%, 0.7\%\}$. The X~axis
represents the size of the exact \textsc{group by} values and the Y~axis, the
corresponding standard error. Both of the X and Y axis are in a logarithmic
scale. The standard error generally decreases when the memory budget increases.
However, for small views, the error can exceed 100\% for Probabilistic Counting
and \textsc{LogLog}: this is caused by a form of overfitting where many
counters are not or barely used (see Section~\ref{sec:probacounting}) when the
ratio of the  view size over the memory budget is  small. In contrast,
Fig.~\ref{fig:expUscensusErrorGibbons} shows that \gt{} has sometimes accuracy
better than 0.01\% for small views. For the multifractal estimation technique
(see Fig.~\ref{fig:expUscensusErrorMulti}), the error decreases when the
sampling ratio increases. While the accuracy can sometimes approach 10\%,
we never have reliable accuracy. 

\subsubsection*{Test over synthetic data} \label{sub:smalldbgen}

\begin{figure*}[!t]
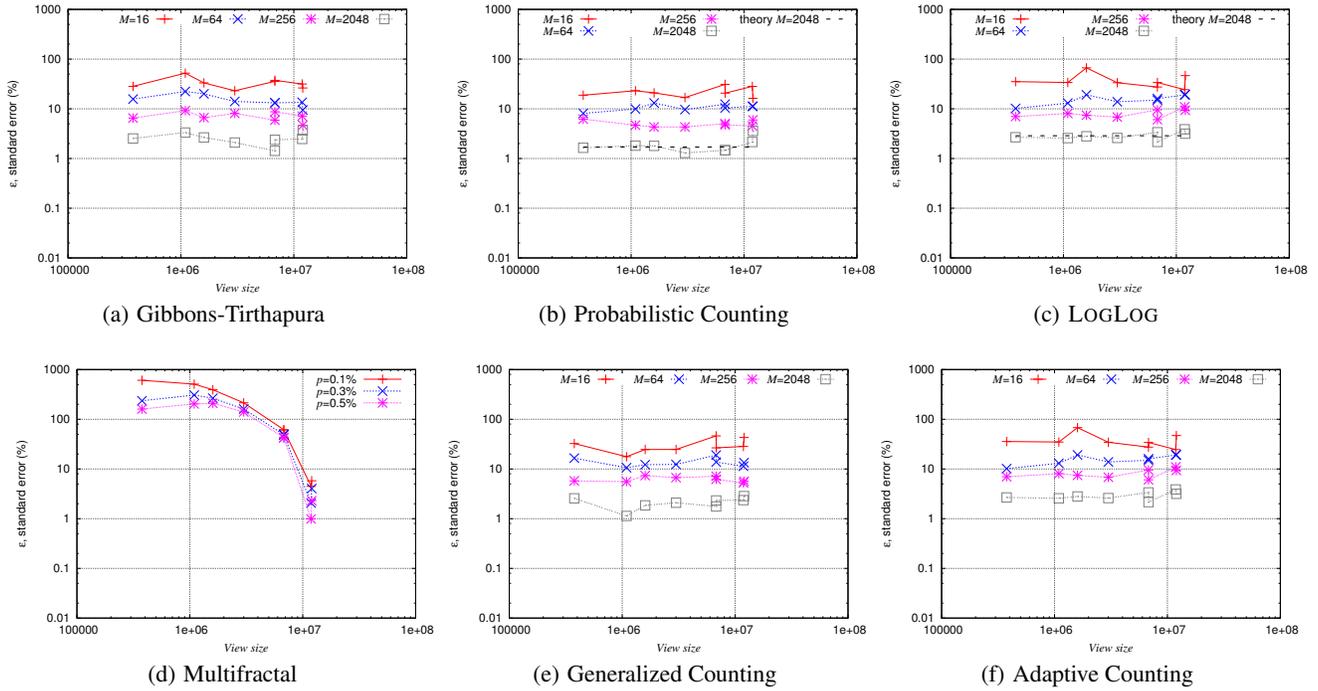

\begin{center}
  \subfigure[{Gibbons-Tirthapura}]{\includegraphics[width=0.32\textwidth]{\myfig{gibbonsdbgen_error}}\label{fig:expDbgenErrorGibbons}}\quad
  \subfigure[Probabilistic Counting]{\includegraphics[width=0.32\textwidth]{\myfig{countingdbgen_error}}\label{fig:expDbgenErrorCounting}}
  \subfigure[\textsc{LogLog}]{\includegraphics[width=0.32\textwidth]{\myfig{loglogdbgen_error}}\label{fig:expDbgenErrorLogLog}}\quad
  \subfigure[Multifractal]{\includegraphics[width=0.32\textwidth]{\myfig{multifractaldbgen_error}}\label{fig:expDbgenErrorMulti}}
  \subfigure[Generalized Counting]{\includegraphics[width=0.32\textwidth]{\myfig{baryossefdbgen_error}}\label{fig:expDbgenErrorBar}}
  \subfigure[Adaptive Counting]{\includegraphics[width=0.32\textwidth]{\myfig{adaptivedbgen_error}}\label{fig:expDbgenErrorAdaptive}}
\end{center}
\caption{Standard error of estimation as a function of exact view size for
increasing values of $M$ (synthetic data set).} \label{fig:expDbgenError}
\end{figure*}

Similarly, we plotted the standard error for each technique, computed from the
DBGEN data set (see Fig.~\ref{fig:expDbgenError}). The five
unassuming techniques have the same behaviour observed on the US Census data
set. The model-based  multifractal technique
(see Fig.~\ref{fig:expDbgenErrorMulti}) is especially accurate
because  DBGEN follows a uniform
distribution~\cite{DBGEN}. For this reason, DBGEN is a poor tool to benchmark model-based
view-estimation techniques, but this problem does not carry over to
 unassuming techniques since they are
data-distribution oblivious.

We also performed experiments on large data sets (5, 10, 20 and 30\,GiB)
generated by DBGEN.\@ Table~\ref{tab:errorcountinggibbons} shows that the
accuracy is not sensitive to data and view sizes for small $M$. In addition,
for large views, Probabilistic Counting has a small edge in accuracy.

\begin{table}[!h]
\caption{\label{tab:errorcountinggibbons}Standard error over large data sets.}
\centering
\begin{scriptsize}
\subtable[\label{tab:errorcounting}Probabilistic Counting]{
\begin{tabular}{|cc|ccc|}\cline{3-5}
     \multicolumn{2}{c}{} & \multicolumn{3}{|c|} {\textbf{Memory budget}} \\  \hline
     \textbf{Data size} & \textbf{View size} & 64   &  128   & 256 \\ \hline
   5\,GiB  & 1000000 & 11\% &   8\%  & 5\% \\
   10\,GiB & 2000000 & 10\% & 7\% & 6\% \\
   20\,GiB & 4000000 & 8\% & 6\% & 5\% \\
   30\,GiB & 6000000 & 9\% & 7\% & 7\% \\
   \hline
\end{tabular}
}
\subtable[\label{tab:errorgibbons}{Gibbons-Tirthapura}]{
\begin{tabular}{|cc|ccc|}\cline{3-5}
     \multicolumn{2}{c}{} & \multicolumn{3}{|c|} {\textbf{Memory budget}} \\  \hline
     \textbf{Data size} & \textbf{View size} & 64   &  128   & 256 \\ \hline
    5\,GiB & 1000000 & 10\% & 8\% & 7\% \\
    10\,GiB & 2000000 & 9\% & 7\% & 6\% \\
    20\,GiB & 4000000 & 10\% & 8\% & 6\% \\
    30\,GiB & 6000000 & 14\% & 8\% & 5\% \\
   \hline
\end{tabular}
} \end{scriptsize}
\end{table}

\subsubsection{Speed} \label{sub:smallmemory}

The time needed to estimate the size of all the views by the unassuming
techniques is about 5~minutes for the US~Census~1990 data set and 7~minutes for
the synthetic data set. For the multifractal technique, all the estimates are
completed in roughly 2~seconds, but  it takes 1~minute (resp. 4~minutes) to
sample 0.5\% of the US Census data set (resp. the synthetic data set -- TPC~H),
in part because the data is not stored in a flat file. We ran further
experiments on the data generated by DBGEN (with a scale factor equal to 5,
i.e., 5\,GiB of data) to highlight the time spent by each processing step:
loading and parsing the data, hashing and computing estimated view sizes. As
shown in Table~\ref{tab:runinngtime}, the running time of the algorithms is
sensitive to the number of dimensions. For a low (resp. high) number of
dimensions, relatively more time is spent reading data (resp. hashing data).
However, the time spent hashing or reading is in turn much larger than the rest
of the time spent by the algorithms (counting). This  explains why all the
unassuming estimation algorithms have similar running times and why timings are
not sensitive to the memory parameter ($M$), as long as it is small.

\begin{table}
    \caption{\label{tab:runinngtime}Wall-clock running times.}
    \centering
    \begin{scriptsize}

    \subtable[\label{tab:runinngtime1}Unidimensional view ($\text{view size}=7.5 \times 10^5$)]{
    \begin{tabular}{|cc|cccccccc|} \cline{3-10}
     \multicolumn{2}{c}{} & \multicolumn{2}{|c}{\textbf{Loading}} &  \multicolumn{2}{c}{\textbf{Hashing}} &  \multicolumn{2}{c}{\textbf{Counting}}
     & \multicolumn{2}{c|}{\textbf{Time (s)}} \\ \hline
     \multicolumn{2}{|c|}{\textbf{Memory}}  &  $m_1$ & $m_2$ & $m_1$ & $m_2$ & $m_1$ & $m_2$ & $m_1$ & $m_2$ \\ \hline
     \multirow{5}*{\rotatebox{90}{\textbf{Technique}}}
     & (1) & 50\% & 52\% & 42\% & 45\% & 7\% & 3\% & 72 & 68\\
     & (2) & 54\% & 40\% & 45\% & 35\% & 1\% & 26\% & 68 & 90 \\
     & (3) & 53\% & 52\% & 46\% & 45\% & 1\% & 3\% & 67 & 68 \\
     & (4) & 54\% & 17\% & 46\% & 14\% & -- & 69\% & 67 & 215 \\
     & (5) & 54\% & 20\% &  46\% & 18\% &  -- & 62\% & 68 & 175 \\
    \hline
    \end{tabular}

    }
    \subtable[\label{tab:runinngtime2}tridimensional view ($\text{view size} = 2.4 \times 10^7 $)]{
     \begin{tabular}{|cc|cccccccc|} \cline{3-10}
     \multicolumn{2}{c}{} & \multicolumn{2}{|c}{\textbf{Loading}} &  \multicolumn{2}{c}{\textbf{Hashing}} &  \multicolumn{2}{c}{\textbf{Counting}}
     & \multicolumn{2}{c|}{\textbf{Time (s)}} \\ \hline
     \multicolumn{2}{|c|}{\textbf{Memory}}  &  $m_1$ & $m_2$ & $m_1$ & $m_2$ & $m_1$ & $m_2$ & $m_1$ & $m_2$ \\ \hline
     \multirow{5}*{\rotatebox{90}{\textbf{Technique}}}
     & (1) & 29\% &  15\% & 68\% & 13\% & 3\% &  72\% & 239 & 240 \\
     & (2) & 30\% & 13\% & 70\% & 11\% & 1\% & 76\% & 235 & 277 \\
     & (3) & 29\% & 15\% & 71\% & 13\% & -- & 72\% & 237 & 240 \\
     & (4) & 30\% & 5\% & 70\% & 5\% & 1\% & 90\% & 235 & 652\\
     & (5) & 29\% & 6\% & 71\% & 5\% & -- & 88\%& 238 & 576 \\
    \hline
    \end{tabular}
    }

    \subtable{
    \begin{tabular}{ccc}
    \multicolumn{3}{c}{$m_1=256$  \hfil $m_2=8388608$} \\
    (1): \textsc{LogLog} & (2): Probabilistic Counting & (3): Adaptive Counting  \\
    \multicolumn{3}{c}{(4): {Gibbons-Tirthapura} \hfil (5): Generalized Counting }  \\
    \end{tabular}
    }
    \end{scriptsize}
\end{table}

\subsection{Large Memory Budgets}\label{sublarge}

When the memory budget is close to the view size,
estimation techniques are not warranted.
Hence, we did not use the US Census data set since it is
too small.

\subsubsection{Accuracy}

\begin{figure*}[t]
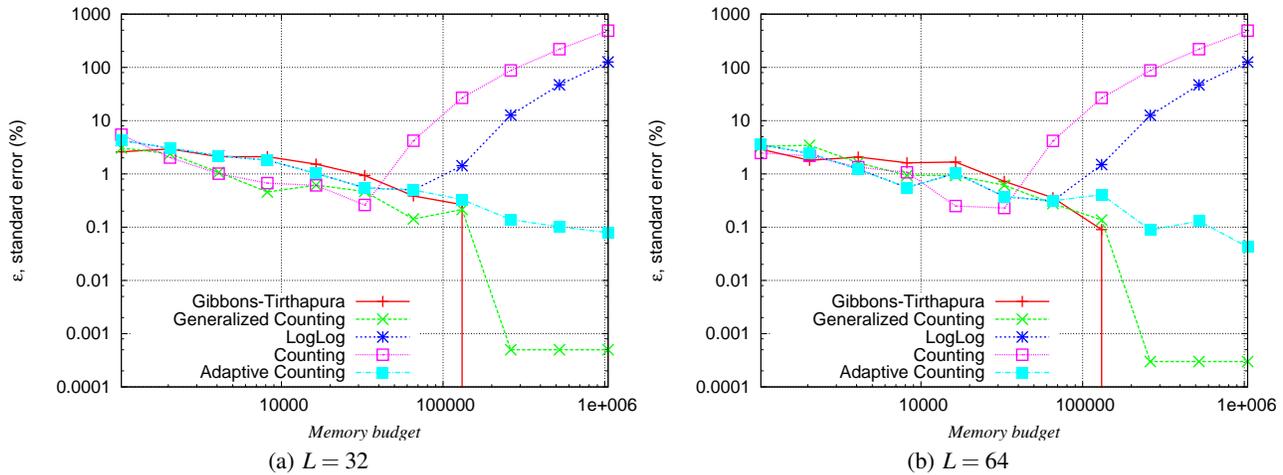

    \subfigure[$L=32$]{\includegraphics[width=0.48\textwidth,angle=0]{\myfig{smallerrorvsmemory32}}}%
    \subfigure[$L=64$]{\includegraphics[width=0.48\textwidth,angle=0]{\myfig{smallerrorvsmemory}}}
    \caption{\label{smallviewerrorvsmemory}Standard error accuracy for a small unidimensional view (250,000 items)  as a function of memory budgets $M$.}
\end{figure*}

\begin{figure*}[t]
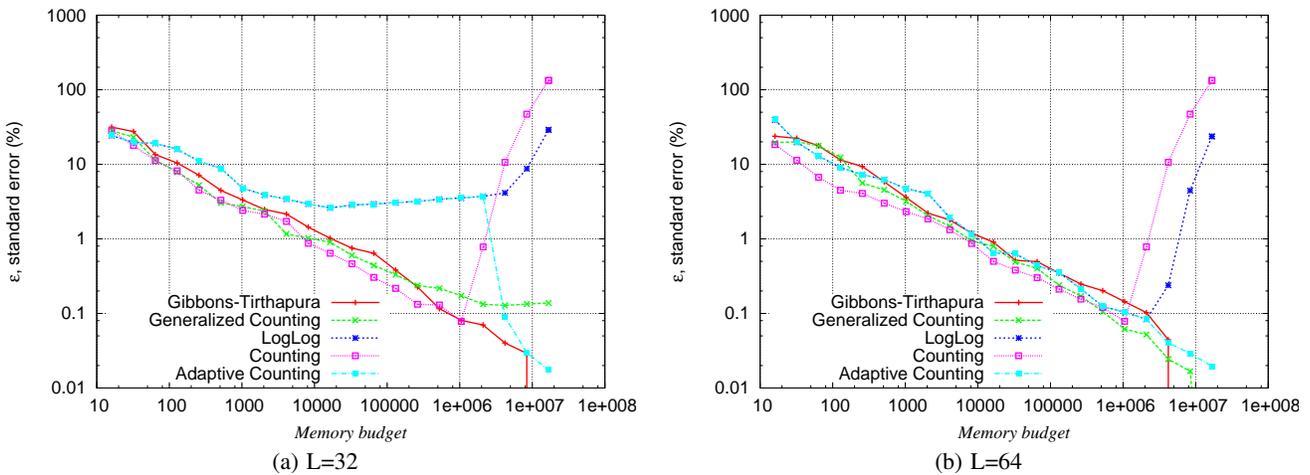

\begin{center}
    \subfigure[L=32]{\includegraphics[width=0.48\textwidth]{\myfig{errorvsmemory_dbgen_l32}}\label{fig:expDbgenErrorGibbons32}}\quad
    \subfigure[L=64]{\includegraphics[width=0.48\textwidth]{\myfig{errorvsmemory_dbgen_l64}}\label{fig:expDbgenErrorCounting32}}
\end{center}
\caption{ \label{fig:expDbgenErrorVSMemoryHM}Standard error of estimation for a
given view (four dimensions and $1.18\times10^7$ distinct tuples) as a function
of memory budgets $M$ (synthetic data set).}
\end{figure*}

Fig.~\ref{smallviewerrorvsmemory} shows the behavior of the five probabilistic
schemes over a moderately small synthetic unidimensional view. While all five
schemes have similar accuracy when the memory budget is small relative to the
size of the view, as soon as the memory budget is within an order of magnitude
of the view size, they differ significantly: \textsc{LogLog} and Counting are
no longer reliable whereas the three other schemes quickly achieve nearly exact
estimates. As we increase the memory budget, this phenomenon happens somewhat
later with \textsc{LogLog} than Counting. Adaptive Counting still has a good
accuracy for large $M$ because it switches from \textsc{LogLog} estimates to
linear counting estimates~\cite{whang1990lin} (see
Algorithm~\ref{algo:loglog}). The accuracy of \gc{} is limited by the size of
$L$. Finally, we ran some tests over a large view using large values of $M$
(see Fig.~\ref{fig:expDbgenErrorVSMemoryHM}): these values of $M$ still
translate in memory usages well below 1\,GiB.  The main difference with the
large view being that \textsc{LogLog} and Adaptive Counting performance seems
to be substantially worst than Probabilistic Counting unless we increase the
number of bits ($L=64$).

\subsubsection{Speed}

\begin{figure}[!h]
\begin{center}
    \includegraphics[width=1\columnwidth]{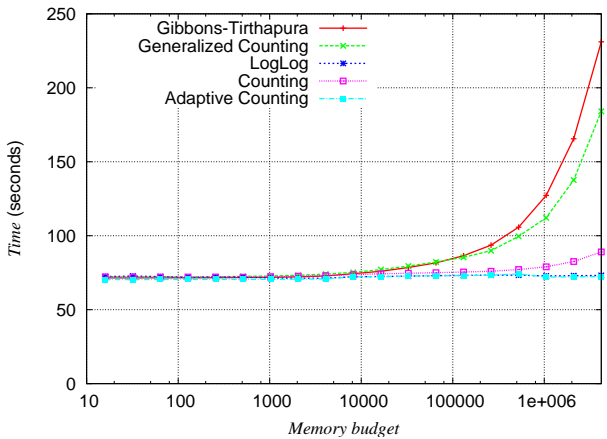}
\end{center}
\caption{\label{fig:expDbgenTimeVSMemoryHM}Estimation time for a given view
(four dimensions and $1.18\times10^7$ distinct tuples) as a function of memory
budgets $M$ (synthetic data set).}
\end{figure}

We also computed the time required to estimate a large view using various
memory budgets $M$ (see Fig.~\ref{fig:expDbgenTimeVSMemoryHM}). For small
values of $M$ ($M\leq 65536$) all techniques are equally fast: most processing
time is spent hashing and parsing the data (see
Table~\ref{tab:errorcountinggibbons}). For larger values of $M$, the time spent
counting the hash values by \gc{} and \gt{} eventually dominates the processing
time (see Table~\ref{tab:runinngtime}). Probabilistic Counting scales well with
large values of $M$ whereas \textsc{LogLog} does not slow down  with increasing
values of $M$, but their accuracies do not necessarily improve either. Adaptive
Counting remains fast and gets increasingly accurate as $M$ becomes large.

\section{Discussion}
Our results show that Probabilistic Counting and \textsc{LogLog} do not
entirely live up to their theoretical promise. For small view sizes relative to
the available memory, the accuracy can be very low. One implication of this
effect is that we cannot increase the accuracy of Probabilistic Counting and
\textsc{LogLog} by adding more memory unless we are certain that all view sizes
are very large. Meanwhile, we observed that \gc{}, \gt{}, and Adaptive Counting
accuracies are independent of the view size and improve when more memory is
allocated, though they also become slower, except for Adaptive Counting which
remains constantly fast. When comparing the memory usage of the various
techniques, we have to keep in mind that the memory parameter $M$ can translate
in different memory usage. The memory usage depends also on the number of
dimensions of each view. Generally, \gc{} and \gt{} will use more memory for
the same value of $M$ than either Probabilistic Counting, Adaptive Counting, or
\textsc{LogLog}, though all of these can be small compared to the memory usage
of the look-up tables $T_i$ used for 3-wise independent hashing. When memory
usage is not a concern ($M$ and $L$ large), \gc{},  \gt{}, and Adaptive
Counting have  accuracies better than 0.1\%. For large values of $M$, which of
\gt{} and \gc{} is more accurate depends on the number of hashing bits used
($L$). \gc{} is the only scheme guaranteed to converge to the true view size as
$M$ grows. View-size estimation by sampling can take minutes when data is not
laid out in a flat file or indexed, but the time required for an unassuming
estimation is even higher. For small values of $M$, streaming and hashing the
tuples accounts for most of the processing time so for faster estimates, we
could store all hashed values in a bitmap (one per dimension).

\section{Conclusion and future work}\label{sec:Conclusion}

We have provided unassuming techniques for view-size estimation
in a data warehousing context. We adapted distinct count estimators to the
view-size estimation problem. 
Using the standard error, we have demonstrated
that among these techniques, \gc{}, \gt{}, and Adaptive Counting provide stable
estimates irrespective of the size of views and that increasing the memory
usage leads to more accuracy.
For small memory budgets, all unassuming methods have comparable speeds.
For large memory budgets, however, only Adaptive Counting
remains constantly fast.  For large view sizes,
using more hashing bits ($L=64$) is important, particularly when using Adaptive Counting.

There is ample room for future work. Firstly, we plan to extend these
techniques to other types of aggregated views (for example, views including
\textsc{having} clauses including icebergs~\cite{iceberg98}). Secondly, we want
to precompute the hashed values for  fast view-size estimation. Furthermore,
these techniques should be tested in a materialized view selection
heuristic~\cite{adbis06ajd}.

\section{Acknowledgements}
The authors wish to thank Owen Kaser from UNB for his contribution to the
software and manuscript as well as  Robert Godin from UQAM for his comments.
This work is supported by NSERC grant 261437 and by  FQRNT grant 112381.
\balance
\bibliographystyle{abbrv}
\bibliography{../bib/lemur}

\begin{thebibliography}{10}

\bibitem{alon1986}
N.~Alon, L.~Babai, and A.~Itai.
\newblock A fast and simple randomized parallel algorithm for the maximal
  independent set problem.
\newblock {\em J. Algorithms}, 7(4):567--583, 1986.

\bibitem{237823}
N.~Alon, Y.~Matias, and M.~Szegedy.
\newblock The space complexity of approximating the frequency moments.
\newblock In {\em STOC '96}, pages 20--29, 1996.

\bibitem{adbis06ajd}
K.~Aouiche, P.~Jouve, and J.~Darmont.
\newblock Clustering-based materialized view selection in data warehouses.
\newblock In {\em ADBIS'06}, volume 4152 of {\em LNCS}, pages 81--95, 2006.

\bibitem{aouicheiceis2007}
K.~Aouiche and D.~Lemire.
\newblock Unassuming view-size estimation techniques in {OLAP}.
\newblock In {\em ICEIS'07}, pages 145--150, 2007.

\bibitem{BarYossef2002}
Z.~Bar-Yossef, T.~S. Jayram, R.~Kumar, D.~Sivakumar, and L.~Trevisan.
\newblock Counting distinct elements in a data stream.
\newblock In {\em RANDOM'02}, pages 1--10, 2002.

\bibitem{cai2005fat}
M.~Cai, J.~Pan, Y.-K. Kwok, and K.~Hwang.
\newblock Fast and accurate traffic matrix measurement using adaptive
  cardinality counting.
\newblock In {\em {MineNet}'05}, pages 205--206, 2005.

\bibitem{durand2003lcl}
M.~Durand and P.~Flajolet.
\newblock Loglog counting of large cardinalities.
\newblock In {\em ESA'03}, volume 2832 of {\em LNCS}, pages 605--617, 2003.

\bibitem{faloutsos1996msd}
C.~Faloutsos, Y.~Matias, and A.~Silberschatz.
\newblock Modeling skewed distribution using multifractals and the 80-20 law.
\newblock In {\em VLDB'96}, pages 307--317, 1996.

\bibitem{iceberg98}
M.~Fang, N.~Shivakumar, H.~Garcia-Molina, R.~Motwani, and J.~D. Ullman.
\newblock Computing iceberg queries efficiently.
\newblock In {\em VLDB'98}, pages 299--310, 1998.

\bibitem{flajolet1985pca}
P.~Flajolet and G.~Martin.
\newblock Probabilistic counting algorithms for data base applications.
\newblock {\em Journal of Computer and System Sciences}, 31(2):182--209, 1985.

\bibitem{Gibbons2001}
P.~B. Gibbons and S.~Tirthapura.
\newblock Estimating simple functions on the union of data streams.
\newblock In {\em SPAA'01}, pages 281--291, 2001.

\bibitem{gol98met}
M.~Golfarelli and S.~Rizzi.
\newblock A methodological framework for data warehouse design.
\newblock In {\em {DOLAP'98}}, pages 3--9, 1998.

\bibitem{graycube}
J.~Gray, A.~Bosworth, A.~Layman, and H.~Pirahesh.
\newblock Data cube: A relational aggregation operator generalizing group-by,
  cross-tab, and sub-total.
\newblock In {\em ICDE '96}, pages 152--159, 1996.

\bibitem{gup97sel}
H.~Gupta.
\newblock Selection of views to materialize in a data warehouse.
\newblock In {\em {ICDT'97}}, pages 98--112, 1997.

\bibitem{haas1995sbe}
P.~Haas, J.~Naughton, S.~Seshadri, and L.~Stokes.
\newblock Sampling-based estimation of the number of distinct values of an
  attribute.
\newblock In {\em VLDB'95}, pages 311--322, 1995.

\bibitem{KDDRepository}
S.~Hettich and S.~D. Bay.
\newblock The {UCI} {KDD} archive.
\newblock \url{http://kdd.ics.uci.edu}, last checked on 23/10/2006, 2000.

\bibitem{viewsizetechreport}
D.~Lemire and O.~Kaser.
\newblock One-pass, one-hash n-gram count estimation.
\newblock Technical Report TR-06-001, Dept. of CSAS, UNBSJ, 2006.
\newblock available from \url{http://arxiv.org/abs/cs.DB/0610010}.

\bibitem{schmidt1993chb}
J.~Schmidt, A.~Siegel, and A.~Srinivasan.
\newblock {Chernoff-Hoeffding} bounds for applications with limited
  independence.
\newblock In {\em SODA'93}, pages 331--340, 1993.

\bibitem{DBGEN}
{TPC}.
\newblock {DBGEN} 2.4.0.
\newblock \url{http://www.tpc.org/tpch/}, last checked on 23/10/2006, 2006.

\bibitem{whang1990lin}
K.-Y. Whang, B.~T. Vander-Zanden, and H.~M. Taylor.
\newblock A linear-time probabilistic counting algorithm for database
  applications.
\newblock {\em ACM Trans. Database Syst.}, 15(2):208--229, 1990.

\end{thebibliography}
\end{document}